%% file: wot_tops_main.tex
\documentclass[sigconf]{acmart}
\usepackage[ruled]{algorithm2e} % For algorithms

\PassOptionsToPackage{hyphens}{url}
\usepackage{subfig}

\usepackage{soul}

\copyrightyear{2022}
\acmYear{2022} 
\setcopyright{iw3c2w3}
\acmConference[CDNG'22]{Proceedings of the CDNG'22}{February 21--22, 2022}{Brisbane, Australia}
\acmBooktitle{Proceedings of the Cyber Defence Next Generation Technology and Science Conference (CDNG), February 21--22, 2022, Brisbane, Australia}
\acmPrice{}
\setcopyright{none}

\author{Muhammad Ikram}
\affiliation{%
  \institution{Macquarie University}
  }
\email{muhammad.ikram@mq.edu.au}

\author{Rahat Masood}
\affiliation{%
  \institution{UNSW}
}
\email{rahat.masood@unsw.edu.au}
\author{Gareth Tyson}
\affiliation{%
 \institution{Queen Mary University of London}}
 \email{g.tyson@qmul.ac.uk}

\author{Mohamed Ali Kaafar}
\affiliation{%
  \institution{Macquarie University}}
\email{dali.kaafar@mq.edu.au}

\author{Roya Ensafi}
\affiliation{%
  \institution{University of Michigan}
}
\email{ensafi@umich.edu}

\begin{document}

\title{A Study of Third-party Resources Loading on Web}

\renewcommand{\shortauthors}{Ikram, et al.}

\begin{abstract}

This paper performs a large-scale study of dependency chains in the web, to find that around 50\% of first-party websites render content that they did not directly load.
Although the majority (84.91\%) of websites have short dependency chains (below 3 levels), we find websites with dependency chains exceeding 30.
Using VirusTotal, we show that 1.2\% of these third-parties are classified as suspicious --- although seemingly small, this limited set of suspicious third-parties have remarkable reach into the wider ecosystem. We find that 73\% of websites under-study load resources from suspicious third-parties, and 24.8\% of first-party webpages contain at least three third-parties classified as suspicious in their dependency chain. By running sandboxed experiments, we observe a range of activities with the majority of suspicious JavaScript codes downloading malware.

\end{abstract}

\maketitle

\section{Introduction}
\label{sec:intro}
\input{files/introduction.tex}

\balance

\normalsize
\bibliographystyle{ACM-Reference-Format}
\bibliography{files/wot_tops_main}

\end{document}

%% file: files/introduction.tex
In the modern web ecosystem, websites often load resources from a range of third-party domains such as ad providers, tracking services and analytics services. 
This is a well known design decision that establishes an \textit{explicit trust} between websites and the domains providing such services. However, often overlooked is the fact that these third-parties can further load resources from other domains, creating a \emph{dependency chain}. This results in a form of \textit{implicit trust} between first-party websites and any domains loaded further down the chain. 

 Consider the \texttt{bbc.com} webpage, an hypothetical example, which loads JavaScript program from  \texttt{widgets.com}, which, upon execution loads additional content from another third-party, say \texttt{ads.com}. Here, \texttt{bbc.com} as the first-party website, \emph{explicitly} trusts \texttt{widgets.com}, but \emph{implicitly}  trusts \texttt{ads.com}. This can be represented as a simple dependency chain in which \texttt{widgets.com} is at level 1 and \texttt{ads.com} is at level 2. Past work tends to ignore this, instead collapsing these levels into a single set of third-parties~\cite{falahrastegar2014anatomy}. 

In this work, we study the dependency chains in the web ecosystem by focusing on the implicitly loaded resources. \section{Dependency Dataset}
First, we obtain the resource dependencies of the Alexa top-200K websites' main pages using the method described in~\cite{ikram2019chain}. 
This Chromium-based Headless~\cite{gHeadless} crawler renders a given website and tracks resource dependencies by recording network requests sent to third-party domains. The requests are then used to reconstruct the dependency chains between each first-party website and its third-party URLs. Note that each first-party can trigger the creation of multiple dependency chains (to form a tree structure). From the Alexa Top-200k websites, we collect 11,287,230 URLs which consist of 6,806,494 unique external resources that correspond to 68,828 and 196,940, respectively, unique second level domains of third- and first-parties.

The \emph{next} challenge is to classify domains as suspicious vs. innocuous. For this we use VirusTotal~\cite{VirusTotal} --- an online solution which aggregates the scanning capabilities provided by 68 Anti-Virus (AV) tools, scanning engines and datasets. It has been commonly used in the academic literature to detect malicious apps, executables, software and domains~\cite{zhao2019decade, Ikram:2016, ikram2017first}. For each domain, we use the VirusTotal \textsf{report} API to obtain the VTscore for each third-party domain. This VTscore represents the number of AV tools that flagged the website as malicious (max.\ 68). 
The reports also contain meta-information such as the first scan date, scan history, and domain category. 
We further supplement each domain with their WebSense~\cite{websense} category provided by the VirusTotal's reports. 
\begin{figure}%[th]
\centering
\subfloat[]{
\includegraphics[width=0.52\columnwidth]{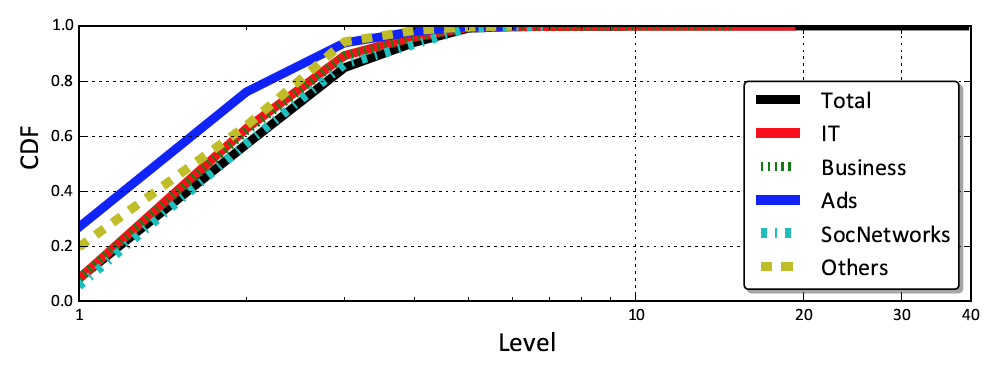}\label{fig:fp_levels}
}
\subfloat[]{
\includegraphics[width=0.46\columnwidth]{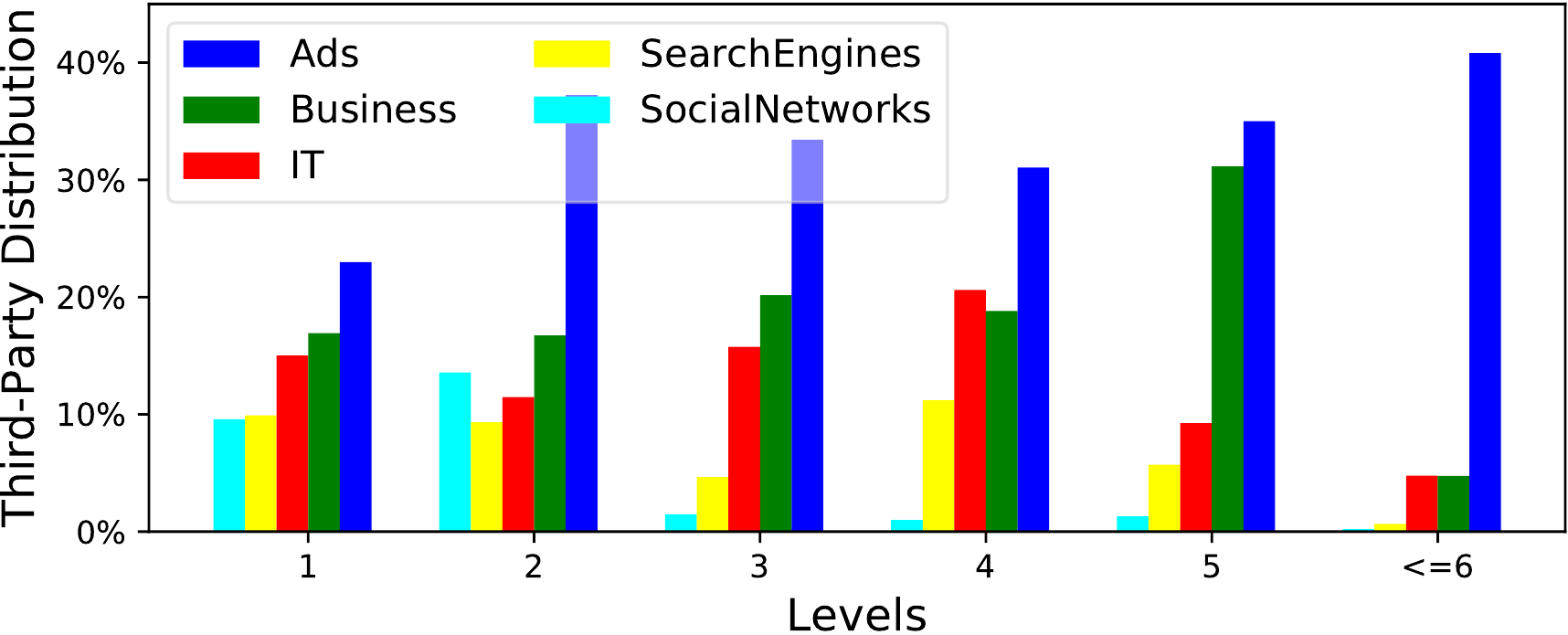}\label{fig:tp_levels}
}
\vspace{-0.3cm}
\caption{\textbf{(a) CDF of dependency chain lengths; and (b) distribution of third-party websites across categories and levels.}} 
\vspace{-0.6cm}
\end{figure}
\section{Exploring the Chains}
We begin by inspecting how extensive dependency chains are across the Alexa's top-200K. We confirm their prominence, finding that around 50\% of websites \emph{do} include third-parties (e.g., content delivery networks (CDNs) such as \texttt{akamaihd.net} and ad and tracking services such as \texttt{google-analytics.com}) which subsequently load other third-parties to form a dependency chain (i.e., they implicitly trust third-parties they do not directly load). The most common \emph{implicitly} trusted third-parties are well known operators, e.g.,  \texttt{google-analytics.com} and \texttt{doubleclick.net}: these are implicitly imported by 68.3\% (134,510) and 46.4\% (91,380) websites respectively. However, we also observe a wide range of more obtuse third-parties such as {\tt pippio.com} and {\tt 51.la} imported by 0.52\% (1,146) and 0.51\% (1,009) of websites. 
As depicted in Figure~\ref{fig:fp_levels}, although the majority (84.91\%) of websites have short chains (with levels of dependencies below 3), we find first-party websites with dependency chains exceeding 30 in length. This not only complicates page rendering, but also creates notable attack surface.

We also inspect the \emph{categories} of third-party domains hosting these resources. Figure~\ref{fig:tp_levels} presents the make-up of third-party categories at each level in the chain. It is clear that, across all levels, advertisement domains make up the bulk of third-parties. We also notice other highly demanded third-party categories such as search engines, Business and IT. 
Figure~\ref{fig:tp_levels} also reveals that the distributions of categories vary across each dependency level. For example, 23.1\% of all loaded resources at level 1 come from advertisement domains, 37.3\% at level 2, and 46.2\% at level 3. In other words, the proportion increases across dependency levels. In contrast, social network third-parties (e.g., Facebook) are mostly presented at level 1 (9.58\%) and 2 (13.57\%) with a significant drop at level 3. The dominance of advertisements is not, however, caused by a plethora of ad domains: there are far fewer ad domains than business or IT. Instead, it is driven by the large number of requests to advertisements: Even though ad domains only make-up 1.5\% of third-parties, they generate 25\% of resource requests. Importantly, these popular providers can trigger further dependencies; for example, \texttt{doubleclick.com} imports 16\% of its resources from further implicitly trusted third-party websites. This makes such domains an ideal propagator of malicious resources for any other domains having implicit trust in it~\cite{ Lauinger2017}.
\vspace{-0.1cm}
\section{Finding Suspicious Chains}
With the above in mind, we then proceed to inspect if \emph{suspicious} or even potentially \emph{malicious} third-parties are loaded via these long dependency chains.
We do not limit this to just traditional malware, but also include third-parties that are known to mishandle user data and risk privacy leaks~\cite{masood2018incognito, ipccc2018, hashmi2019longitudinal}. 

Using the VirusTotal service API, we classify third-party domains into innocuous vs. suspicious. When using a classification threshold (i.e., VTscore $\geq$ 10, we find that 1.2\% of third-parties are classified as suspicious. 
Although seemingly small, we find that this limited set of suspicious third-parties have remarkable reach. 73\% of websites under-study load resources from suspicious third-parties, and 24.8\% of first-party webpages contain at least 3 third-parties classified as suspicious in their dependency chain. This, of course, is impacted by many considerations which we explore --- most notably, the power-law distribution of third-party popularity, which sees a few major players on a large fraction of websites. 

%\vspace{-0.1cm}
\vspace{-0.08cm}
\section{Characterizing Suspicious JavaScript Resources}
In our past research~\cite{ikram2019chain}, we inspected the prevalence of dependency chains in the web. Here, we build on these past findings to focus on \emph{what} activities are undertaken within the dependency chains. Hence, we \textit{sandbox} all suspicious JavaScript programs to monitor their activities. We build a sandbox and perform tests executing suspicious JavaScript codes.
We find that JavaScript codes loaded at higher levels in the dependency chain ($ Level \geq$2) generated a larger number of HTTP requests. This is worrying as resources loaded at higher levels in the dependency chain are the most opaque to the website operator (i.e., they rely on implicit trust). 
The activities of these scripts are diverse. For example, we find evidence of first-party websites performing malicious search poisoning activities when (implicitly) loading some JavaScript codes. The most typical purpose of the suspicious JavaScript code is downloading dropfiles. Dropfiles are executables (e.g., malware, Exploitkits, Trojans, etc) exploiting the browser to download and execute code without user consent. We also observe instances of \emph{very} active JavaScript codes, e.g., the most active (at level 4) downloads 129 files. 

It is also interesting to observe that the resources at level $\geq$ 2 tend to have higher VTscores, indicating that their activities are blocked by a large number of virus checkers. 
The actual content of the files are quite diverse. 
We exclude the 8\% which are encrypted, and therefore cannot be examined. 
The vast majority of remaining files (98.62\%) are Adware and Click bots, suggesting that these types of financial gain are a major driving force in this domain. The remainder are Potentially Unwanted Programs (0.52\%), Exploitkits (0.36\%), Adware and Click Bots (98.62\%), and Trojan (0.50\%).
For instance, \url{videowood.tv/assets/js/poph.js} uses and exploits {\tt eval()} --- JavaScript's dynamic loading method --- to download and execute {\tt 1832-fc204a9bcefeab3d.exe} (with VTscore=5). This then enables the attacker to take over web browser for displaying a wide range of adverts and garner fraudulent clicks.

\vspace{-0.08cm}
\section{Conclusion}
Inspired by the lack of prior work focusing on how resources are loaded, we explored dependency chains in the web ecosystem and found that over 50\% of websites \emph{do} rely on implicit trust. 
Although the majority (84.91\%) of websites have short chains (with levels of dependencies below 3), we found first-party websites with chains exceeding 30 levels. The most common \emph{implicitly} trusted third-parties are well known operators (e.g., \texttt{doubleclick.net}), but we also observed various less known implicit third-parties.
We hypothesised that this might create notable attack surfaces. 
To confirm this, we classified the third-parties using VirusTotal to find that 1.2\% of third-parties are classified as potentially malicious. Worryingly, our ``confidence" in the classification actually increases for implicitly trusted resources (i.e., trust level $\geq$ 2), where 78\% of suspicious JavaScript resources have a VTscore $>52$. In other words, more implicitly trusted JavaScript resources have high VTscores than explicitly trusted ones.
These resources have remarkable reach --- largely driven by the presence of highly central third-parties, e.g., \texttt{google-analytics.com}.
With this in mind, we performed sandbox experiments on the suspicious JavaScript codes to understand their actions. We witnessed extensive download activities, much of which consist of downloading dropfiles and malware. It was particularly worrying to see that JavaScript resources loaded at  level $\geq$ 2 in the dependency chain tended to have more aggressive properties, particularly as exhibited by their higher VTscore. 
This exposes the need to tighten the loose control over indirect resource loading and implicit trust: it creates exposure to risks such as malware distribution, search engine optimization (SEO) poisoning, malvertising and exploit kit redirection. We argue that ameliorating this can only be achieved through transparency mechanisms that allow web developers to better understand the resources on their webpages (and the related risks). To foster future research, we share all our datasets, experimental testbed code and scripts with the wider research community: {\tt https://wot19submission.github.io}.